# Quantitative Wavefront sensing with static Foucault and pyramid tests

François Hénault[1], Yan Feng[2], Alain Spang[3], Laura Schreiber[4]

[1] Optical Engineering Consulting, 740 Chemin d'Argevillières, 07000 Veyras – France
[2] LIRA, Observatoire de Paris, Université PSL, Université Paris Cité, Sorbonne Université, CY Cergy Paris Université, CNRS, 75014 Paris – France
[3] Université Côte d'Azur, Observatoire de la Côte d'Azur, CNRS, Laboratoire Lagrange – France
[4] INAF - Osservatorio di Astrosica e Scienza dello Spazio di Bologna
Via Gobetti 93/3, 40129 Bologna – Italy

E-mail: henaultfbm@gmail.com

## ABSTRACT

Wavefront sensors (WFS) are now core components in the fields of metrology of optical systems, biomedical optics and adaptive optics systems for astronomy. Nowadays, the most popular WFS is the Shack-Hartmann, which is fully static but suffers from a limited spatial resolution in the pupil plane of the tested optical system. Higher spatial resolutions are achievable with other types of sensors, e.g. the pyramid WFS that requires temporal modulation of the recorded signals and implies high mechanical and electronic complexity. This paper examines the possibility of performing quantitative wavefront sensing inspired from the well-known Foucault test and only comprising static, non-modulated optical components. Here, two candidate designs of static WFS are proposed, based on a set of reflective prisms. Those prisms may be coated with gradient density filters. A simplified mathematical model allows for the definition of the wavefront slopes reconstruction formula and for the calculation of the wavefront itself. Numerical simulations demonstrate that the wavefront measurement accuracy is compliant with classical diffraction limit criteria when using coated prisms. Thus accurate WFE measurements are feasible in that case.

Keywords: Wavefront sensing; Wavefront error, Optical metrology, Biomedical optics, Adaptive optics, Fourier optics

## 1 INTRODUCTION

Wavefront sensors (WFS) have now become core components in the fields of metrology of optical systems [1], biomedical optics [2], and Adaptive optics (AO) systems for astronomy [3]. They play a major role in terms of Wavefront error (WFE) measurement accuracy, spatial resolution inside the exit pupil of the considered optical systems, and capture range of their WFE, therefore enabling to assess the ultimate achievable performance. From a practical point of view however, these systems should also be static, i.e. including no moving components of any kind such a translation and rotation stages or tip-tilt mirrors, especially when rapidly time-varying WFEs such as those generated by atmospheric disturbance above a ground-based astronomical telescope have to be measured. Establishing a list of the different WFS designs used so far and comparing them together is beyond the scope of the present paper, but it is useful and somewhat inspiring to summarize schematically a few decisive stages of their development throughout the years.

Historically, the first of these WFS is indeed the Foucault knife-edge test [4], allowing the observation of black and white images of the exit pupil of the tested optics, sometimes named "Foucaultgrams". Such images are closely related to the slopes of the WFE to be measured. Because it only requires standard and inexpensive accessories, the Foucault test is widely spread in the field of instrument control and remains very popular among amateur astronomers. However, it is reputed for its limited ability to provide quantitative and accurate data because the produced Foucaultgrams are barely readable by experienced operators and only reveal low spatial frequency WFE aberrations. Following the development of modern computers, Wilson [5] proposed a direct inversion process between the Foucaultgrams and the WFE transmitted by the optical system. He demonstrated that such a relationship effectively exists but is only applicable to weak amplitude errors, thus considerably limiting the practical range of application of the method. Later, it was demonstrated that quantitative WFE measurements are achievable with the Foucault test when the knife-edge is displaced laterally in the image plane along two orthogonal axes [6-7], then modulating the recorded electrical signals. Hence, the method requires a dynamic modulation of the knife-edge filter. Although this system proved to be very efficient for measuring stationary aberrations, it must be concluded that quantitative WFE measurements cannot be carried out by use of a static Foucault WFS.

A significant breakthrough occurred during the 1980' with the development of the Shack-Hartmann (SH) wavefront sensor [8]. This device is fully static (i.e. no moving parts of any kind) and achieves quantitative WFE measurements with fair accuracy. Its principle consists in slicing the exit pupil of the tested optics by means of a grid of equally-spaced micro-lenses, and recording their individual images in the focal plane of the sensor. There small displacements of the sub-images with respect to their nominal locations can be observed and are directly proportional to the WFE slopes errors, from which the WFE itself can be reconstructed. Nowadays, the SH-WFS probably is the most popular one because of its simplicity in realization, limited encumbrance in complicated optical setups, and relatively low price. Actually its main drawback resides in a limited spatial resolution on the exit pupil of the tested optical system, which is governed by the total number of micro-lenses (typically 50 x 50 at most) and makes it insensitive to high spatial frequency WFE aberrations.

One should also mention the Optical differentiation sensor (ODS) where the classical Foucault knife-edge is replaced with a Gradient transmission filter (GTF) that delivers quantitative and accurate measurement results. In 1972, Sprague and Thompson [9] defined a linear ramp of amplitude transmittance having the basic property of optically differentiating the objects under observation. However, their work was essentially focused on improving the contrast of phase-objects in the field of microscopy. Later, a similar principle was applied to the WFE measurements of optical systems. The first WFS explicitly based on a GTF was described by Bortz [10] and later studied by other authors [11-13] in view of its application to AO systems. Experimental studies were previously conducted and demonstrated good measurement accuracy with high spatial resolution on the tested objects [14-15]. However the ODS cannot be considered as being entirely static, since the WFE slopes have to be measured along two orthogonal directions at least, thus implying rotation movements. Finally, it has to be noted that two fully static variants in this WFS family have recently been described, but require having four different off-axis illuminating sources at least [16-17].

Each of the previous test systems shows some drawbacks. In particular, the Foucault and ODS WFS require implementing a fast mobile component located into the image plane of the tested optics to modulate the signals, generally with a high temporal frequency. On the other hand, the fully static SH-WFS suffers from its limited spatial resolution. This paper aims at gathering the best characteristics of all of them, therefore searching for a device achieving simultaneously quantitative and accurate WFE measurement, a high spatial resolution over the pupil of the tested optics, and avoiding any modulation need, in other words. being fully static. For that purpose the historical Foucault test is revisited in section 2, showing two options under the form of bi-prisms. It includes tentative optical designs, theoretical analysis, and a set of numerical simulations evaluating the achievable measurement accuracy of such WFS. A similar approach and numerical evaluations are carried out for the case of a static pyramidal test in section 3. Finally, a brief conclusion is drawn in section 4.

## 2  STATIC FOUCAULT TEST

This section first describes the general layout of the static Foucault WFS (§ 2.1). Then, a family of prisms suitable for integration into the WFS is defined (§ 2.2). A theoretical analysis is presented in § 2.3, including the WFE slopes reconstruction process (§ 2.4). Numerical simulations are carried out in § 2.5 for performance assessment.

### 2.1  General layout and scientific notations

Figure 1 shows a schematic diagram of the static Foucault WFS, where the red and blue lines indicate the field and pupil rays, respectively. The employed coordinate systems are as follows.
- The OXYZ reference frame is attached to the exit pupil of the tested optical system. The point O is the pupil centre, and OZ is its optical axis. Points P in the OXY pupil plane are denoted by their Cartesian coordinates (x,y), and the WFE to be measured is noted as Δ(x,y).
- The O'X'Y'Z reference frame is attached to the image plane of the tested optical system. Point O' is its focal point at a distance $f$ = OO' along the optical axis. Points M' in the O'X'Y' plane are denoted by their Cartesian coordinates (x',y'). The Foucault knife-edge filter is installed here, shown as a two-side reflective prism with vertex at O'.
- The O"X"Y"Z reference frames are symmetrical one with respect to the other, and attached to the images of the pupil planes onto the detector arrays. Points P" located in the O"X"Y" planes are denoted by their Cartesian coordinates (x",y"). Because the OXY and O"X"Y" planes are optically conjugated via the lenses L1+ and L1- and assuming a magnification factor equal to unity, the (x,y) and (x",y") coordinates are considered equivalent in a first-order approximation.

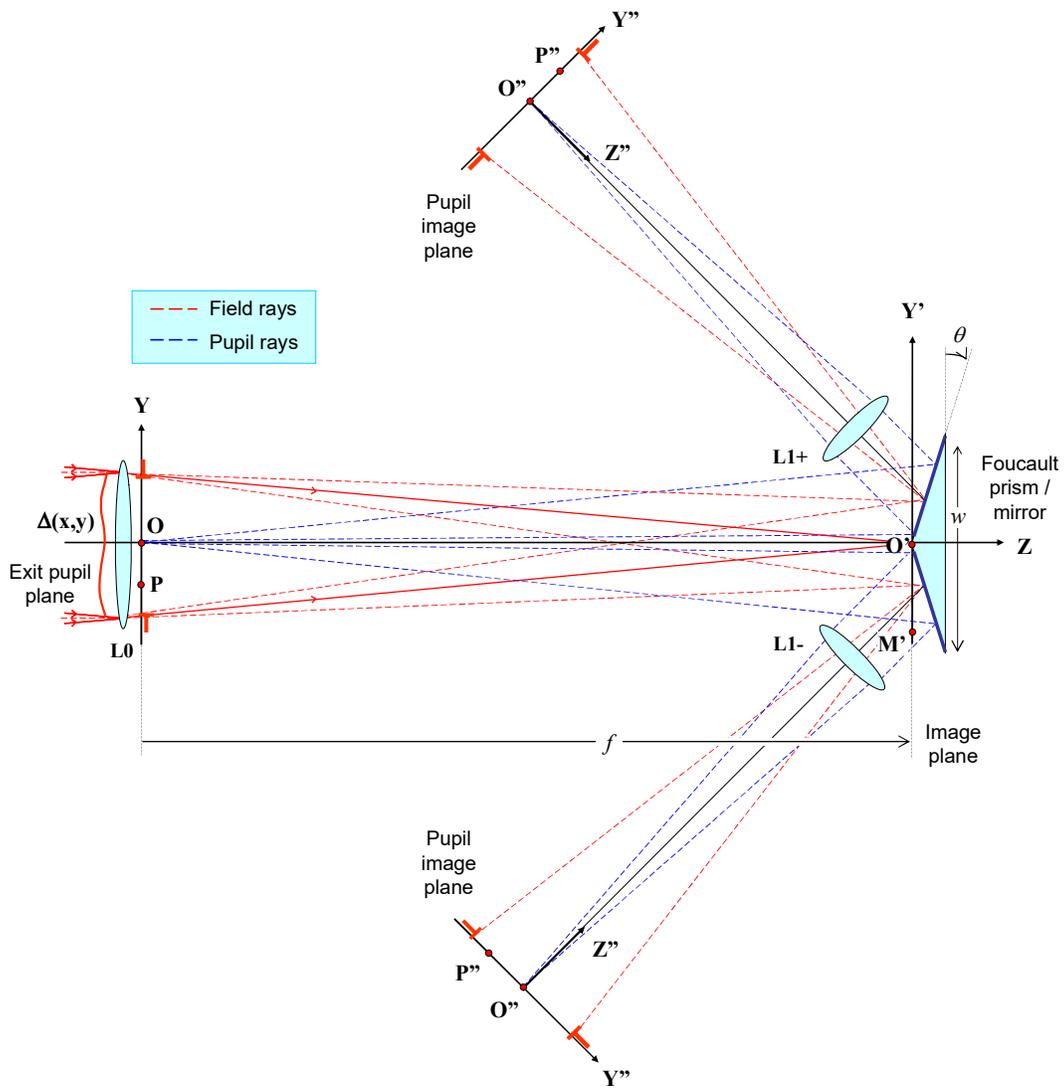

Figure 1: Schematic diagram of the static Foucault WFS. The X, X' and X" axes are perpendicular to the plane of the sheet. Note the presence of circular pupil image stops in the O"X"Y" planes.

In addition the following parameters and scientific notations used in the text are indicated in Table 1.

Table 1: Parameters and scientific notations.

| Parameter | Symbol | Value | Unit |
|---|---|---|---|
| Wavelength of the incoming electromagnetic radiation | $\lambda$ | 0.5 | micron |
| Wavenumber of incoming electromagnetic radiation | $k = 2\pi/\lambda$ | 125663,7 | $cm^{-1}$ |
| Exit pupil diameter of the tested optical system | $D$ | 0.5 | m |
| Focal length of the tested optical system | $f = OO'$ | 10 | m |
| Foucault prism angle | $\theta$ | 2 | deg. |
| Foucault prism width | $w$ | 10 | mm |
| Amplitude transmission of the filter along Y'-axis (GTF) | $t(y')$ | See Eq. 2 | - |

## 2.2 Optical designs

As sketched in Figure 1, the optical design of the Foucault WFS is based on a two-side reflective prism located at or near to the image plane O'X'Y' of the tested optical system, thus allowing simultaneous acquisitions of two symmetrical images on their associated detector arrays. Although the prism in the Figure reflects the impinging rays, the same principles remain applicable to a refractive or semi-reflective prism without loss of generality. Two different options are examined as illustrated in Figure 2. Since the edges of the prism are directed along the X'-axis, their shape and slopes are defined along the Y'-axis. The same principles hold for a prism rotated by 90° by simply permuting the X' and Y' axes.

### 2.2.1 Sharp Foucault prism (SFP)

In the SFP design the reflective prism acts as the classical Foucault knife edge along both the +Y' and –Y' directions. Mathematically, the reflective faces of the prism are defined as:

$$z = y' \tan \theta \quad \text{if } y' \geq 0, \text{ and}$$
$$z = -y' \tan \theta \quad \text{if } y' < 0. \tag{1}$$

This basic prism is illustrated in Figure 2-a and its slopes along the Y'-axis are equal to $\partial z / \partial y' = \pm \tan \theta$. In that case the amplitude transmission $t(y')$ of the filter is uniformly equal to 1.

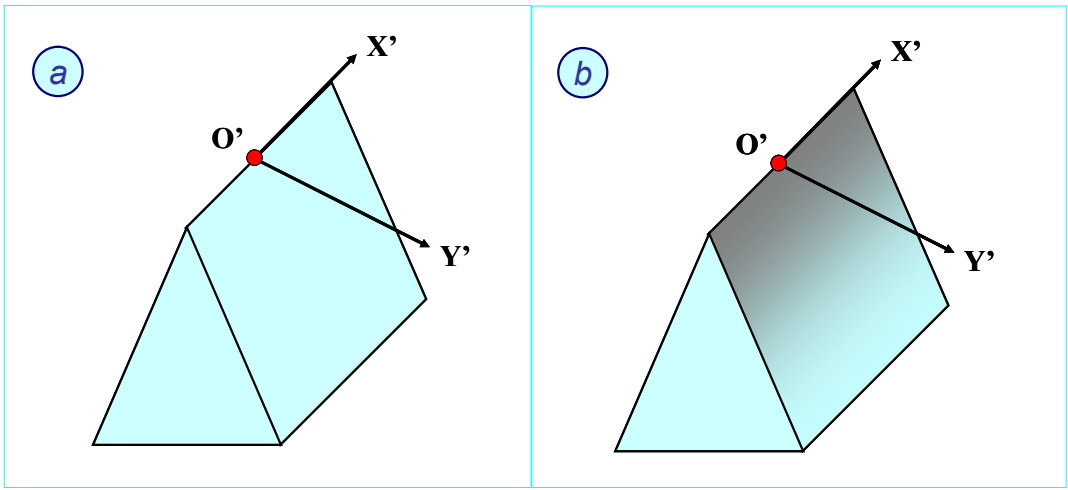

Figure 2: Two different designs of the Foucault filter in the O'X'Y' plane. a) Case of the basic SFP. b) Case of the SDP. See text for details.

### 2.2.2 Sharp differentiating prism (SDP)

Let us now transform the basic SFP into a variant of the optical differentiation WFS, whose basic principle has been described and discussed in detail by Bortz and Oti [10-11]. It consists in adding a linear amplitude transmission filter (i.e. a GTF) into the O'X'Y' image plane to encode the WFE slopes into grey intensity variations observable in the pupil image planes O"X"Y". Practically speaking, this GTF may be coated directly onto the faces of the prism as shown in Figure 2-b. Mathematically, the filter transmission $t(y')$ is written as:

$$t(y') = a\,y'/p + b \qquad \text{if } |y'| \leq p(1-b)/a \text{, and}$$
$$t(y') = 1. \qquad \text{if } |y'| > p(1-b)/a,$$
(2)

where $p$ is proportional to the slope of the transmission filter, and $(a,b)$ is a couple of parameters comprised between 0 and 1 used to balance the "slope" and "offset" factors of the GTF. Figure 3 shows three different curves $t(y')$ with parameter $b = 0.5$ and $a/p$ ratios equal to 0.4, 0.5 and 0.6 respectively.

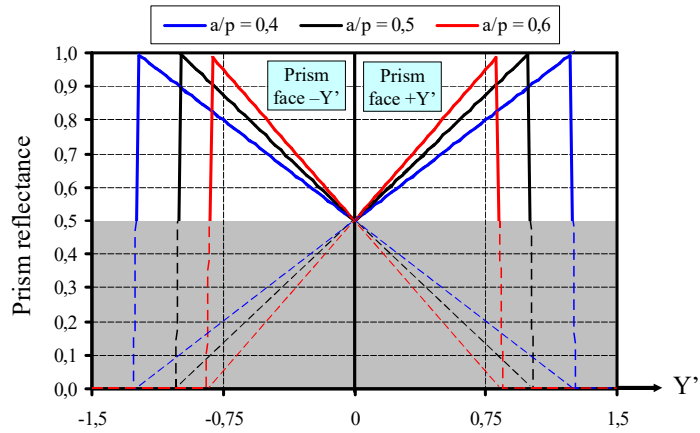

Figure 3: Transmission curves of a SDP with sharp edge located at coordinate y' = 0, parameter $b = 0.5$ and $a/p$ ratios equal to 0.4, 0.5 and 0.6. The dashed curves indicate the transmissions of a classical ODS.

This Figure illustrates the main advantage of the SDP with respect to a classical ODS. Whereas the latter is equipped with a GTF linearly varying from zero to unity (as shown by dashed lines in the figure), the lowest amplitude transmission of the SDP is equal to 50% at the sharp edge of the prism. This naturally results from the optimal choice of parameter $b = 0$, Hence the SPD transmission is always higher than that value, which actually provides a significant radiometric gain when compared with the ODS. In fact, the SDP may be considered as the combination of two symmetric ODS, where only the higher transmission areas of the GTF are reflected or crossed by the incident beam. Then the SDP provides all necessary information in order to reconstructing the WFE slopes along one given direction in a simultaneous, single-shot measurement.

### 2.2.3 Prisms arrangement

The prisms previously described can be combined to perform simultaneous measurements of the WFE slopes along the X" and Y" axes, as schematically illustrated in Figure 4. It implies using one Beamsplitter (BS), a set of Folding mirrors (FM) and ancillary optics (not shown in the Figure) to re-image the sliced pupil images of the tested optics onto the detector arrays. Here, PX' and PY' denote those prisms whose edges are aligned along the X' and Y' axes respectively. This arrangement applies to any of the two previous prisms.

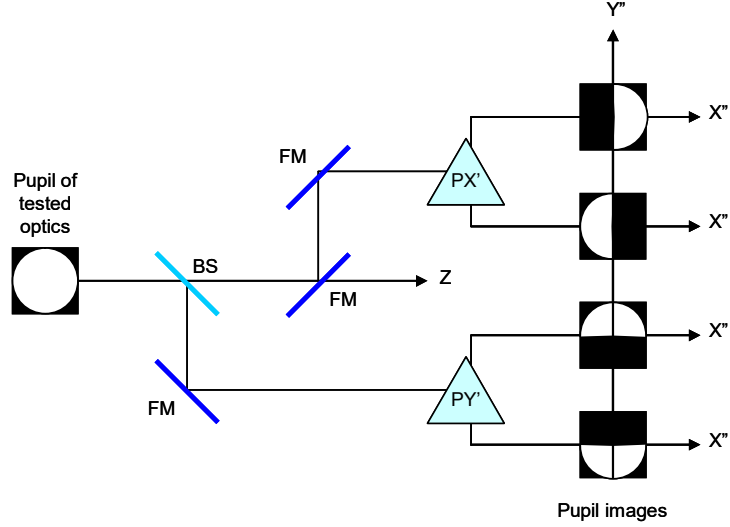

Figure 4: Prisms arrangement usable for simultaneous measurements of the WFE slopes along the X" and Y" axes.

Interestingly, all the proposed solutions are integrated into a static WFS, therefore requiring no modulation of any optical component in the system. It is worth studying how they form the pupil images of the tested optical system and deriving the reconstruction formulas of the WFE slopes. This is the scope of the following subsection.

## 2.3 Theoretical analysis

The mathematical symbols used in this subsection and their physical meaning are summarized in the table below. The magnification factor from the pupil plane of the tested optics to its images formed on the detector array is assumed to be equal to one. Therefore, the Cartesian coordinates (x,y) and (x",y") are considered as strictly equivalent in the following text and in the mathematical notations indicated in Table 2.

Table 2: Full and condensed mathematical notations.

| Symbol | Condensed symbol | Physical meaning | Unit |
| --- | --- | --- | --- |
| $i$ | – | Complex square root of –1 | – |
| $B_D(x,y)$ | $B_D$ | Pupil transmission map in the exit pupil plane OXY of the tested optics. It is defined as a pillbox function equal to unity inside a disk of diameter $D$ and to zero outside of it | – |
| $\Delta(x,y)$ | $\Delta$ | Wavefront error (WFE) to be measured | waves |
| $\partial\Delta(x,y)/\partial x$ | $\Delta'_X$ | WFE slopes along X-axis | rad |
| $\partial\Delta(x,y)/\partial y$ | $\Delta'_Y$ | WFE slopes along Y-axis | rad |
| H(y') | H | Heavyside distribution | |
| δ(y') | δ | Dirac's "delta" distribution | m-1 |
| δ'(y') | δ' | First derivative of Dirac's delta distribution | m-2 |

We write the complex amplitude diffracted from the exit pupil plane of the tested optics into the O'X'Y' filter plane as a direct Fourier transform [18]:

$$A'(x',y') = -\frac{i}{\lambda f} \iint_{x,y} B_D(x,y) \exp[ik\Delta(x,y)] \exp[-ik(xx'+yy')/f] \, dx\, dy, \quad (3)$$

where the $\lambda$ and $f$ parameters are defined in Table 1. Multiplying $A'(x',y')$ with the amplitude transmission function of the filter $t(y')$ in Eq. 2 and returning into the pupil plane via an inverse Fourier transform, leads to an expression of the retro-diffracted complex amplitude $A''_\pm(x'',y'')$:

$$A''_\pm(x'',y'') = \frac{1}{\lambda^2 f^2} \iint_{x',y'} \left( \iint_{x,y} B_D(x,y) \exp[ik\Delta(x,y)] \exp[-ik(xx'+yy')/f] \, dxdy \right) \left( \pm a\frac{y'}{p} + b \right) H(\pm y') \exp[ik(x'x''+y'y'')/f] \, dx'dy' \quad (4)$$

Inverting both integration sums, Eq. 4 rewrites as:

$$A''_\pm(x'',y'') = \frac{1}{\lambda^2 f^2} \iint_{x,y} B_D(x,y) \exp[ik\Delta(x,y)] \left( \iint_{x',y'} \exp[-ik(x'(x-x'')+y'(y-y''))/f] \left( \pm a\frac{y'}{p} + b \right) H(\pm y') \, dx'dy' \right) dxdy \quad (5)$$

or, setting new variables $u = (x-x'')/\lambda f$ and $v = (y-y'')/\lambda f$:

$$A''_\pm(x'',y'') = \iint_{u,v} B_D(x''+\lambda fu, y''+\lambda fv) \exp[ik\Delta(x''+\lambda fu, y''+\lambda fv)] \left[ \iint_{x',y'} \exp[-2i\pi(x'u+y'v)] \left( \pm a\frac{y'}{p} + b \right) H(\pm y') \, dx'dy' \right] dudv \quad (6)$$

Recognizing that the x',y' summation term in Eq. 6 is a Fourier transform, and knowing that the transform of $H(\pm y')$ is equal to $\hat{H}(v) = \delta(v)/2 \mp i/2\pi v$ [19] leads to:

$$A''_\pm(x'',y'') = \iint_{u,v} B_D(x''+\lambda fu, y''+\lambda fv) \exp[ik\Delta(x''+\lambda fu, y''+\lambda fv)] \left[ \pm i\frac{a}{4\pi p} \left( \delta'(v) \pm \frac{i}{\pi v^2} \right) + \frac{b}{2} \left( \delta(v) \mp \frac{i}{\pi v} \right) \right] \times \delta(u) \, dudv \quad (7)$$

Setting $u = 0$ and using classical properties of Dirac's delta derivative $\delta'(v)$, $A''_\pm(x'',y'')$ rewrites as:

$$A''_\pm(x'',y'') = \pm i\frac{a}{4\pi p} \int_v \delta(v) \frac{\partial(B_D(x'',y''+\lambda fv)\exp[ik\Delta(x'',y''+\lambda fv)])}{\partial y''} dv - \frac{a\lambda f}{4\pi^2 p} \int_v \frac{1}{v^2} B_D(x'',y''+\lambda fv) \exp[ik\Delta(x'',y''+\lambda fv)] dv$$
$$+ \frac{b}{2} \int_v \delta(v) B_D(x'',y''+\lambda fv) \exp[ik\Delta(x'',y''+\lambda fv)] dv \pm i\frac{b}{2\pi} \int_v \frac{1}{v} B_D(x'',y''+\lambda fv) \exp[ik\Delta(x'',y''+\lambda fv)] dv \quad (8)$$

Or, changing again variable $\lambda fv = y - y''$ with $dv = dy/\lambda f$:

$$A''_\pm(x'',y'') = \pm i\frac{a}{4\pi p} \int_y \delta(y-y'') \frac{\partial(B_D(x'',y)\exp[ik\Delta(x'',y)])}{\partial y''} dy - \frac{a\lambda f}{4\pi^2 p} \int_y \frac{1}{(y''-y)^2} B_D(x'',y) \exp[ik\Delta(x'',y)] dy$$
$$+ \frac{b}{2} \int_y \delta(y-y'') B_D(x'',y) \exp[ik\Delta(x'',y)] dy \mp i\frac{b}{2\pi} \int_y \frac{1}{y''-y} B_D(x'',y) \exp[ik\Delta(x'',y)] dy \quad (9)$$

Since it was demonstrated in Ref. [6] that the presence of functions $B_D(x'',y'')$ inside the integral sums only generates small diffraction effects at the pupil rim of the tested optical system, they can be removed from the summation terms. Therefore Eq. 9 becomes in condensed notations:

$$A''_\pm(x'',y'') = \underbrace{\pm \frac{af}{2p} B_D \Delta'_Y \exp[ik\Delta] - a\frac{\lambda f}{4\pi^2 p} B_D \left( \exp[ik\Delta] * \frac{1}{y''^2} \right)}_{\text{Optical differentiation terms}} + \underbrace{\frac{b}{2} B_D \exp[ik\Delta] \mp i\frac{b}{2\pi} B_D \left( \exp[ik\Delta] * \frac{1}{y''} \right)}_{\text{Foucault knife-edge terms}} \quad (10)$$

where the Optical differentiation and Foucault knife-edge terms are proportional to the coefficients $a$ and $b$ respectively, and the symbols * denote convolution products.

### 2.4 WFE slopes reconstruction formulas

#### 2.4.1 Foucault terms

As pointed out by Barakat [20], the second component of the Foucault terms $A_{F\pm}(x'',y'')$ in Eq. 10 is a convolution product related to the Cauchy principal value distribution, which has no simple analytical expression. Thus he recommended solving it with the help of numerical simulations. However it is possible to derive a heuristic relation when assuming that the 1/y" function can be approximated as:

$$\frac{1}{y''} \approx -F\lambda \delta'(y''), \quad (11)$$

where $\delta'(y'')$ is the derivative of Dirac's delta distribution and $F$ is an arbitrary, non dimensional coefficient. Denoting them $A''_{F\pm}(x'', y'')$, the Foucault terms in Eq. 10 become:

$$A''_{F\pm}(x'', y'') \approx \frac{b}{2} B_D \exp[ik\Delta] \pm i\frac{b}{2\pi} F\lambda B_D (\exp[ik\Delta] * \delta'(y'')), \quad (12)$$

and owing to the differentiation properties of convolution products, an analytical expression of $A_{F\pm}(x'', y'')$ is found to be:

$$A''_{F\pm}(x'', y'') \approx \frac{b}{2} B_D \exp[ik\Delta] \pm i\frac{b}{2\pi} F\lambda B_D \left(\frac{\partial \exp[ik\Delta]}{\partial y''} * \delta(y'')\right)$$

$$= bB_D \left(\frac{1}{2} \mp F\Delta'_Y\right)\exp[ik\Delta]. \quad (13)$$

### 2.4.2 Optical differentiation terms

From Eq.10 the optical differentiation terms are defined as:

$$A''_{D\pm}(x'', y'') = \pm \frac{af}{2p} B_D \Delta'_Y \exp[ik\Delta] - a\frac{\lambda f}{4\pi^2 p} B_D \left(\exp[ik\Delta] * \frac{1}{y''^2}\right); \quad (14)$$

In opposition to the Foucault terms, they can be expressed as an exact analytical expression, which is demonstrated into the Appendix A. It leads to

$$A''_{D\pm}(x'', y'') = B_D \frac{af}{2p}\left[\pm \Delta'_Y + \frac{i-1}{2\pi}\right]\exp[ik\Delta] \quad (15)$$

### 2.4.3 Final expressions

Inserting the previous expressions of $A''_{F\pm}(x'', y'')$ in Eq. 13 and $A''_{D\pm}(x'', y'')$ in Eq. 15 into the general relation 10 gives the total complex amplitude retro-propagated to the pupil image plane O"X"Y":

$$A''_\pm(x'', y'') = \left[\pm\left(\frac{af}{2p} - bF\right)\Delta'_Y + \frac{af}{4\pi p}(i-1) + \frac{b}{2}\right] B_D \exp[ik\Delta]. \quad (16)$$

Then the intensities recorded from each side of the prism are equal to the square modulus of $A''_\pm(x'', y'')$:

$$I''_\pm(x'', y'') = |A''_\pm(x'', y'')|^2 = B_D^2\left[\left(\frac{af}{2p} - bF\right)^2 \Delta'^2_Y \pm \left(\frac{a^2f^2}{4\pi p^2} + \frac{abf}{2p}\left(1 - \frac{F}{2}\right) - Fb^2\right)\Delta'_Y + \frac{b^2}{4} - \frac{abf}{4\pi p} + \frac{a^2f^2}{8\pi^2 p^2}\right] \quad (17)$$

and their difference is directly proportional to the searched WFE slopes $\partial\Delta(x,y)/\partial y$ along the Y-axis:

$$I''_+(x'', y'') - I''_-(x'', y'') = B_D^2 \left(\frac{a^2f^2}{2\pi p^2} + \frac{abf}{p}\left(1 - \frac{F}{2}\right) - 2Fb^2\right)\frac{\partial\Delta(x,y)}{\partial y} \quad (18)$$

That last relation directly leads to a slopes reconstruction formula along the Y-axis applicable to both the SFP and SDP WFS that is:

$$\frac{\partial\Delta(x,y)}{\partial y} = \frac{I''_+(x'', y'') - I''_-(x'', y'')}{G B_D^2} \quad (19)$$

where $G$ is the "gain factor" of the static WFS equal to:

$$G = \frac{a^2f^2}{2\pi p^2} + \frac{abf}{p}\left(1 - \frac{F}{2}\right) - 2Fb^2 \quad (20)$$

A similar relation holds for reconstructing the WFE slopes $\partial\Delta(x,y)/\partial x$ along the X-axis.

### 2.4.4 Empirical assumption

The gain expression in Eq. 20 results from an exact approximation in the restrictive case of the SDP when its parameter $b$ is equal to zero. However the Foucault terms in Eq. 10 involve an arbitrary parameter $F$ whose value should be determined experimentally. This statement applies to the SFP (§ 2.2.1) and to any SDP case (§ 2.2.2) when its parameter $b$ differs from zero. This pleads in favor of an empirical adjustment of the gain factor $G$. Practically speaking, it entails a pre-calibration sequence of these WFS with respect to a set of previously known WFE aberrations. This calibrating operation has been included in the numerical simulations presented in the next subsection.

### 2.4.5 Throughput estimation

The throughput value is important feature of any WFE sensing device. Here an approximate throughput estimation can be derived from Eq. 16, assuming that the input WFE is null, and then its partial derivative $\Delta'_Y$ can be set to zero. Since $B_D$ is uniformly equal to unity inside the pupil of the tested optics, Eq. 16 simplifies as:

$$A''_\pm(x'',y'') = \frac{1}{2}\left(b - \frac{af}{2\pi p} + i\frac{af}{2\pi p}\right), \quad (21)$$

therefore producing an average intensity equal to:

$$I''_\pm = \frac{1}{4}\left(b^2 - \frac{abf}{\pi p} + \frac{a^2 f^2}{2\pi^2 p^2}\right). \quad (22)$$

Here the multiplying factor 1/4 naturally appears because four different images of the exit pupil are generated by the WFS. Omitting this irreducible coefficient the throughput can be considered as being equal to unity in the cases of the SFP since $a = 0$ and $b = 1$. In the cases of the SDP however, a throughput loss occurs due to the presence of the GTF ($a \neq 0$). Here the quantity $af/\pi p$ will need to be optimized carefully. Low throughput for this sensor would likely disqualify its use in astronomical applications.

## 2.5 Numerical simulations

### 2.5.1 Numerical model

The employed numerical model is divided into seven main steps. Figure 5 schematically summarizes the whole process.

1) Two different types of Wavefront errors $\Delta(x,y)$ are introduced into the numerical model, namely low and mid-order Zernike polynomials, and random phase errors:

   A. Case of low/mid-order Zernike polynomials: this type of WFE is characteristic of optical aberrations and/or mechanical deformation modes of mirrors and lenses. It also represents differential errors between the science and metrology channels inside an AO system. The original WFE is built from randomly defined coefficients of the first 48 Zernike polynomials.

   B. Case of random phase errors: this type of WFE is typical of the atmospheric disturbance experienced during astronomical observations, which may be partly compensated for by an AO system. Here are considered moderate perturbations characterized by a relative spatial coherence radius equal to $r_0/D$, with $r_0$ the coherence radius equal to 0.5 m and the telescope diameter $D = 10$ m.

2) Following Eq. 3, the complex amplitude $A'(x',y')$ diffracted from the exit pupil of the tested optics to the O'X'Y' filter plane is computed with a Fast Fourier transform (FFT).

3) Then $A'(x',y')$ is multiplied with the transmission of the selected prism. Here the two cases are distinguished:

a) For the basic SFP case in § 2.2.1, the transmissions are simply equal to Heavyside distributions $H(\pm y')$ as defined in Eq. 4 with parameters $a$ and $b$ equal to 0 and 1 respectively.

b) For the SDP case in § 2.2.2, Eq. 4 is still applicable, and the Heavyside distributions are multiplied with the transmission function $t(y')$ defined in Eq. 2, where the parameters $a/p$ and $b$ are both equal to 0.5.

This step and the following must be repeated for both orientations of the prisms, i.e. having their edge parallel to the X' and Y' axes.

4) Computing the inverse FFT of the previous results now gives access to the complex amplitudes $A''_\pm(x'', y'')$ retro-diffracted into the pupil image planes O"X"Y".

5) Then the image intensities $I''_\pm(x'', y'')$ recorded by the detector arrays are taken equal to the square modules of $A''_\pm(x'', y'')$.

6) The WFE slopes $\partial\Delta(x,y)/\partial x$ and $\partial\Delta(x,y)/\partial y$ along the X and Y axes are reconstructed from the image intensity differences; according to Eq. 20 and assuming an optimal gain factor $G$ (see § 2.4 for details).

7) Finally, the WFE $\Delta(x,y)$ is reconstructed from its slopes $\partial\Delta(x,y)/\partial x$ and $\partial\Delta(x,y)/\partial y$ by use of an iterative Fourier transforms algorithm [21] that was selected in view of its accuracy and short computation time.

Numerical simulations were then carried out following that procedure, and their results are summarized in the next subsection.

NOTA   The here above numerical model has been cross-checked with the so-called "phase screen" model developed by Carbillet *et al.* [22] in the framework of a pyramid sensor. The results from both numerical models were found to be in very good agreement.

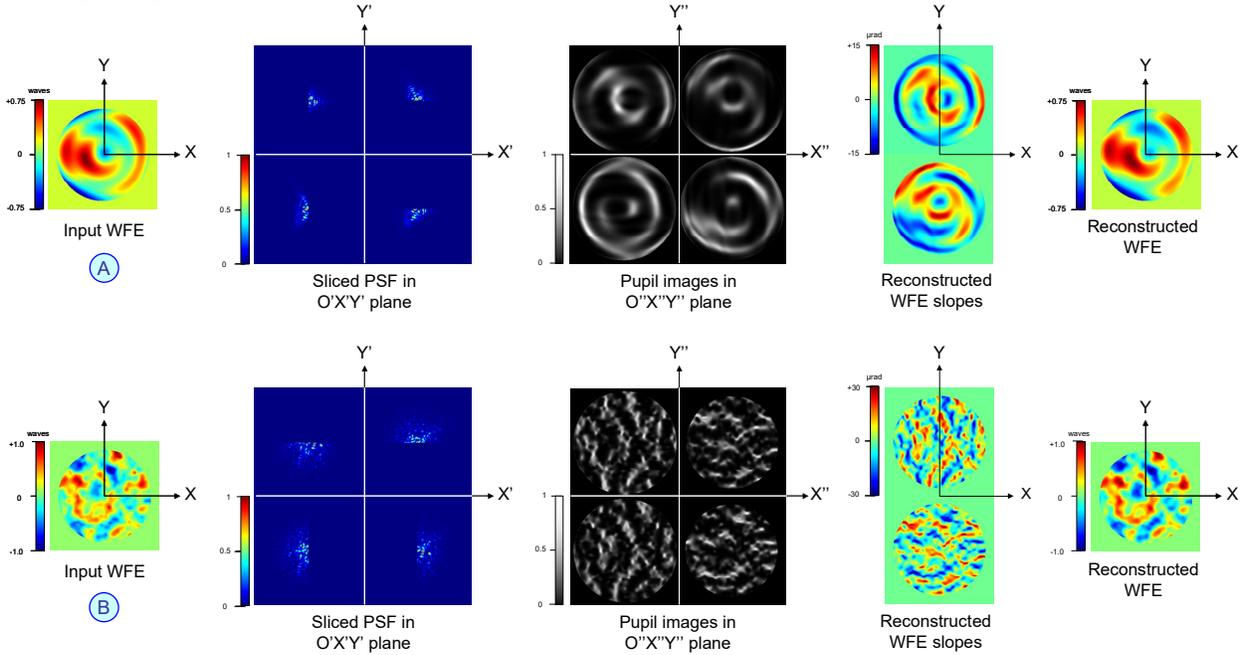

Figure 5: Illustrating the numerical model of the static Foucault test for the cases of low/mid-order Zernike polynomials (A) and random phase errors (B).

*Numerical results*

The results obtained with the previous numerical model are presented as synthetic histograms displayed in Figure 6, showing the achieved measurement accuracy of the WFS in terms of Peak-to-Valley (PTV) and Root mean square (RMS) numbers. Both cases A (low and mid frequency WFE) and B (random phase errors) described in section 2.5.1 have been considered. For each case, the results obtained using the SFP and SDP WFS are illustrated by different histogram colors in the Figure.

The Figure 6 also indicates the diffraction limit criteria, here chosen as:

- <u>Rayleigh's criterion</u>. The PTV measurement error must be lower or equal than $\lambda/4 = 0.25\ \lambda$.
- <u>Maréchal's criterion</u>. The RMS measurement error must be lower or equal than $\lambda/13.5 \approx 0.07\ \lambda$

The reader will find all the detailed results into the Appendix B. In particular, they include PTV and RMS statistics of the reconstructed WFEs and their slopes (see Table B1).

From the results reported in Figure 6 it is apparent that there is a noticeable improvement of the WFE measurement accuracy both in terms of the PTV and RMS criteria when changing the basic SFP into a SDP WFS. Although it does not attain the PTV diffraction limit criterion, especially for the challenging case B simulating random errors, it can be seen that the RMS diffraction limit is overcome in both cases A and B with some margin. It is remarkable that the SDP WFS achieves such performance whatever is the input WFE case. These favorable results may pave the way to simpler WFS designs. Although closed-loop operations are not discussed in this paper, it may be reasonably hoped that better performance will be achieved using the SPD WFS.

In the next section the same principles will be applied to the case of the static pyramid test.

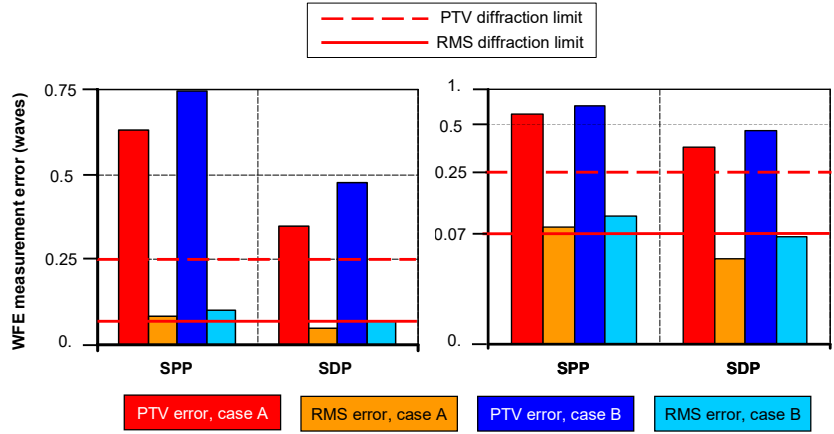

Figure 6: Wavefront measurement errors of the static Foucault test for the cases of low/mid-order Zernike polynomials (case A) and random phase errors (case B). Left side, vertical axis in linear scale. Right side, vertical axis in logarithmic scale.

## 3 STATIC PYRAMID TEST

### 3.1 Principle

In the field of AO for astronomy, a revisited, bi-dimensional version of the Foucault test gave birth to a novel measurement device named pyramidal WFS [23], where the knife-edge is replaced with a four-face glass pyramid. Like for the dynamic Foucault test, a temporal modulation of the pyramid filter must be introduced, moving the filter itself or using a scanning mirror located upstream into the optical system. In both cases, mobile equipment is required and must be operated at a high temporal frequency to compensate for atmospheric perturbations. However, the mechanical and electrical complexity of such components stands for a big challenge, hence the desire to operating with a static pyramid filter. This may be achieved by inserting a diffusing plate near the system's image plane, as described in Ref. [24]. More recent studies also explored ways to improve the data reduction process of a static pyramid, e.g. using optimization and deep learning algorithms [25-27]. Nevertheless, they remain difficult to implement and time-consuming. This may be the reason why the pyramidal test is rarely used in laboratories or industry.

Most of the principles described in section 2 can be reused for the static pyramid test, following the modified prism arrangement depicted in Figure 7. It is equivalent to the single pyramid prism of the original design [23] that allowed minimizing the number of ancillary optics, e.g. fold mirrors, not shown in Figure 7. However this advantage is somewhat counterbalanced by a more complex manufacturing process of the pyramid. The same SDP apodizing process illustrated in Figure 2-b is fully applied to the prisms, either on their bi-prism or pyramid versions.

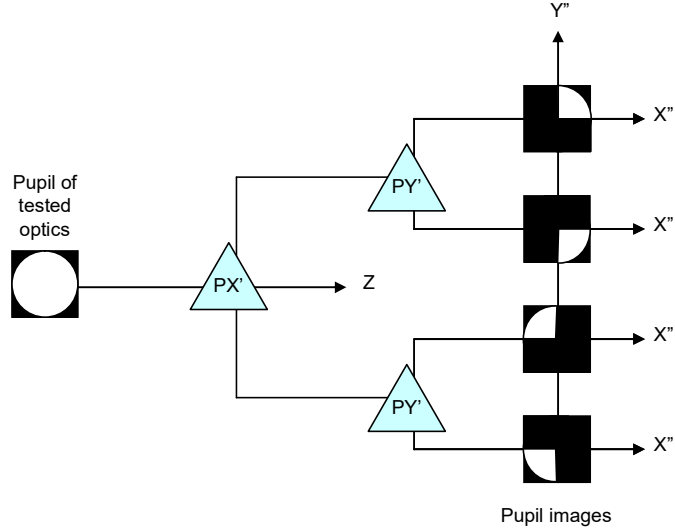

Figure 7: Prisms arrangement usable for the static pyramid test.

## 3.2 Numerical simulations

The numerical model employed for the static pyramid test is quite similar to that described in subsection 2.5.1 and is illustrated in Figure 8. The main differences between them are as follows:

- In step n° 3, the mono-dimensional Heavyside distributions $H(\pm y')$ and their apodizing factors along the Y'-axis are replaced with bi-dimensional products $H(\pm x')H(\pm y')$ with similar multiplying factors directed along both the X' and Y' axes.
- In step n° 6, the slopes reconstruction in Eq. 19 is replaced with the classical relations of the modulated pyramidal WFS:

$$\frac{\partial \Delta(x,y)}{\partial x} = \frac{\bar{I}_{++}(x'',y'') - \bar{I}_{-+}(x'',y'') + \bar{I}_{+-}(x'',y'') - \bar{I}_{--}(x'',y'')}{GB_D^2}$$

$$\frac{\partial \Delta(x,y)}{\partial y} = \frac{\bar{I}_{++}(x'',y'') + \bar{I}_{-+}(x'',y'') - \bar{I}_{+-}(x'',y'') - \bar{I}_{--}(x'',y'')}{GB_D^2}$$

(23)

where $\bar{I}_{++}(x'',y'')$, $\bar{I}_{-+}(x'',y'')$, $\bar{I}_{+-}(x'',y'')$ and $\bar{I}_{--}(x'',y'')$ are the pupil images formed from each quadrant of the pyramid WFS.

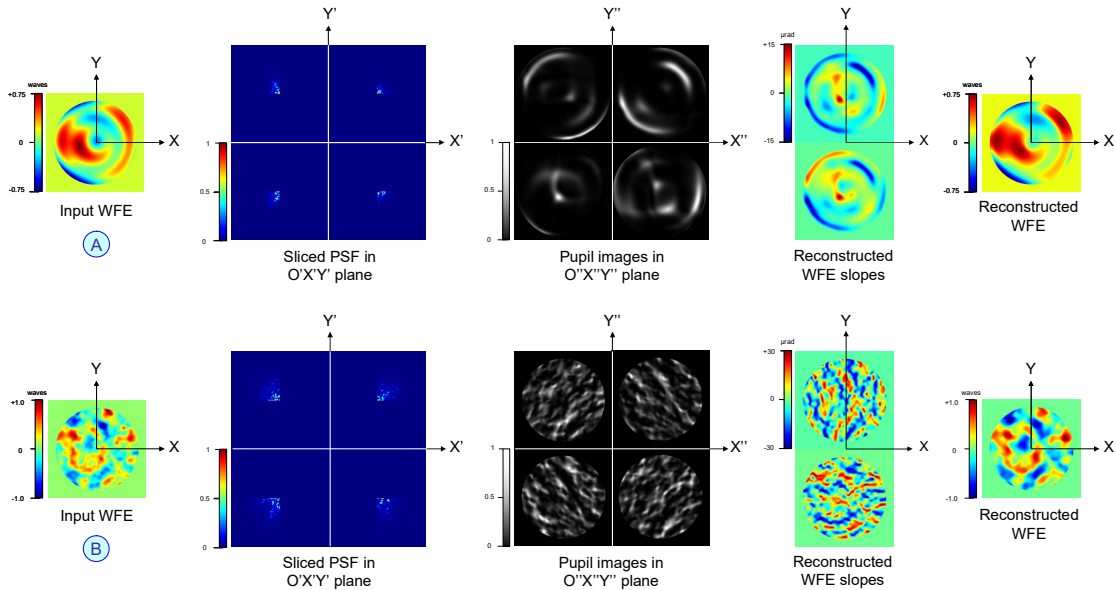

Figure 8: Illustrating the numerical model of the static pyramidal test for the case of low/mid-order Zernike polynomials (case A) ) and random phase errors (B).

The numerical results obtained with this modified numerical model are illustrated by the synthetic histograms displayed in Figure 9. The achieved measurement accuracy of the input WFE in terms of PTV and RMS numbers are shown for the same input WFEs, i.e. low/mid frequency aberrations (case A) and random phase errors (case B). The two different WFS types are considered, integrating either the SFP or SDP. The same diffraction limit criteria as in § 2.5.2 are used, and the detailed results will be found in the Table C1 of Appendix C.

At first glance, the obtained numerical results look similar to those of the static Foucault test. However a more careful examination of the detailed numerical results in Appendix C reveals a small inferiority in terms of the WFE measurement accuracy with respect to the static Foucault test with the SPD, since the RMS diffraction limit is no longer met by about 16 and 18 % in both cases A and B respectively. This is confirmed by the images of the reconstructed slopes and WFE depicted on the fight side of Figure 8, showing irregular patterns that were not visible in Figure 5. This might be due to the necessary multiplication of the complex amplitude with two perpendicular Heavyside distributions in the image plane, while only one occurs in the Foucault configuration. Finally, and despite of the small apparent difference, it might be concluded that the bi-prism configuration depicted in Figure 4 is preferable to the usual pyramid configuration of Figure 7.

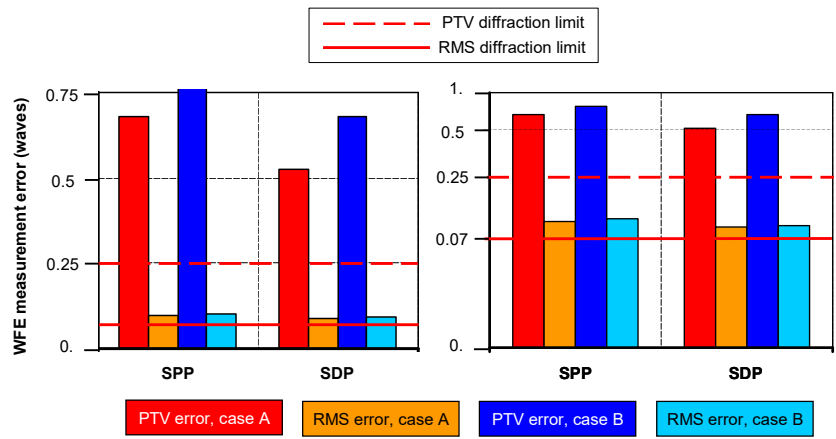

Figure 9: Wavefront measurement errors of the static pyramid test for the cases of low/mid-order Zernike polynomials (case A) and random phase errors (case B). Left side, vertical axis in linear scale. Right side, vertical axis in logarithmic scale.

## 4  CONCLUSION

This paper evaluated the possibility of performing quantitative Wavefront sensing based on the well-known Foucault and pyramid tests, comprising only static, non-modulated optical components. After summarizing the historical background, two candidate designs of Wavefront sensors (WFS) were proposed. They are all based on reflective bi-prisms, standard or customary components. In particular, their faces may be coated with a gradient density filter, like an optical differentiation WFS does.

After a detailed description of these prisms, we carried out a simplified mathematical analysis to define the WFE slopes reconstruction formula of the WFS. A numerical model based on physical optics was then developed to assess the WFE measurement accuracy of the sensors, which was compared to classical diffraction limit criteria. The results of these numerical simulations showed that precise WFE measurements are feasible in the case when the sharp-edge Foucault prism is replaced with a coated differentiating prism (SDP). They also demonstrated that although their performance looks similar, the static pyramid test is a little less accurate than its Foucault counterpart. Finally, compared with the currently developed WFS, the SDP design may pave the way to a much simpler and cheaper static, motion-free WFS generation.

Some important features, such as the Signal-to-Noise Ratio (SNR), capture range, linearity of these WFS and their ability to cope with Non-common path aberrations (NCPA) are not covered in the present paper. However these notions are more relevant in the field of adaptive optics for astronomy, and they deserve future work. It is also worth validating the manufacturing process of the SDP, and integrating it into a complete WFS prototype.

## APENDIX A. COMPUTING OPTICAL DIFFERENTIATION TERMS

From Eq. 14 in the main text the general expression of the optical differentiation terms $A_{D\pm}(x'',y'')$ writes as

$$A''_{D\pm}(x'',y'') = \pm \frac{af}{2p} B_D \Delta'_Y \exp[ik\Delta] - a\frac{\lambda f}{4\pi^2 p} B_D \left( \exp[ik\Delta] * \frac{1}{y''^2} \right) \quad (A1)$$

Its second term is proportional to the convolution product $\exp[ik\Delta] * 1/y''^2$, noted $A_2(x'',y'')$ in the following. That last quantity can be determined by means of a double Fourier transformation:

1) Firstly, the Fourier transform (*FT*) of $A_2(x'',y'')$ writes as:

$$\hat{A}_2(u,v) = FT(\exp[ik\Delta]) \times FT(1/y''^2) = -2\pi^2 v\, sign(v) \times FT(\exp[ik\Delta]), \quad (A2)$$

where $sign(u)$ is the sign function equal to $H(u) - H(-u)$. Hence:

$$\hat{A}_2(u,v) = FT(\exp[ik\Delta]) \times FT(1/y''^2) = -2\pi^2 (vH(v) - vH(-v)) \times FT(\exp[ik\Delta]). \quad (A3)$$

2) Secondly, let us returning to $A_2(x'',y'')$ that is the inverse Fourier transform (*FT*$^{-1}$) of $\hat{A}_2(u,v)$, now writing as:

$$A_2(x'',y'') = -\frac{2\pi}{\lambda}\left[FT^{-1}(uH(u)) + FT^{-1}(uH(-u))\right] * \exp[ik\Delta]. \quad (A4)$$

Using the differentiation and symmetry properties of Fourier transformations, Eq. A4 yields:

$$A_2(x'',y'') = -\frac{2\pi^2(i-1)}{\lambda}\delta(x'',y'') * \exp[ik\Delta] = -\frac{2\pi^2(i-1)}{\lambda}\exp[ik\Delta]. \quad (A5)$$

3) Finally inserting Eq. A5 into Eq. A1 the expression of $A''_{D\pm}(x'',y'')$ turns to be:

$$A''_{D\pm}(x'',y'') = B_D \frac{af}{2p}\left[\pm \Delta'_Y + \frac{i-1}{2\pi}\right]\exp[ik\Delta] \quad (A6)$$

which demonstrates Eq. 15 in the main text.

# APPENDIX B. STATIC FOUCAULT TEST – DETAILED NUMERICAL RESULTS

In this Appendix are presented detailed data tables supporting the numerical simulations of the static Foucault test discussed in subsection 2.5.2. They include:

- The cases of both studied prisms, namely the SFP and SDP.
- The cases of both considered input Wavefront errors, i.e. low and mid frequency WFE (case A) and random phase errors (case B). Here the factors fwfe indicate the sign of the WFE and d an eventual added defocus.
- For each case are indicated the Peak-to-Valley (PTV) and RMS statistics of the wavefronts and their slopes along the X and Y axes, including their reference and measured values, the differences between them, and relative measurement errors in terms of %.

Table A1: Detailed numerical results obtained for the static Foucault test.

### SFP - Case A

| Zern48 | fwwe = 1 | | a = 0 | b = 1 | d = 0 |
|---|---|---|---|---|---|
| Error type | Reference | Measured | Difference | Relative Difference (%) | |
| X-slopes (mrad) | 0,0308 | 0,0258 | 0,0187 | 61 | PV |
| | 0,0061 | 0,0061 | 0,0024 | 39 | RMS |
| Y-slopes (mrad) | 0,0308 | 0,0258 | 0,0187 | 61 | PV |
| | 0,0061 | 0,0061 | 0,0024 | 39 | RMS |
| Wavefront Error (waves) | 1,413 | 0,980 | 0,632 | 45 | PV |
| | 0,261 | 0,248 | 0,083 | 32 | RMS |

### SFP – Case B

| Random | fwwe = 1 | | a = 0 | b = 1 | d = 0 |
|---|---|---|---|---|---|
| Error type | Reference | Measured | Difference | Relative Difference (%) | |
| X-slopes (mrad) | 0,0580 | 0,0411 | 0,0398 | 69 | PV |
| | 0,0080 | 0,0080 | 0,0039 | 48 | RMS |
| Y-slopes (mrad) | 0,0651 | 0,0453 | 0,0493 | 76 | PV |
| | 0,0082 | 0,0082 | 0,0041 | 50 | RMS |
| Wavefront Error (waves) | 2,122 | 1,692 | 0,745 | 35 | PV |
| | 0,331 | 0,315 | 0,101 | 31 | RMS |

### SDP – Case A

| Zern48 | fwwe = 1 | p = 2 µm | a=0,5 | b=0,5 | d = 0 |
|---|---|---|---|---|---|
| Error type | Reference | Measured | Difference | Relative Difference (%) | |
| X-slopes (mrad) | 0,031 | 0,026 | 0,011 | 37 | PV |
| | 0,006 | 0,006 | 0,002 | 27 | RMS |
| Y-slopes (mrad) | 0,031 | 0,026 | 0,011 | 37 | PV |
| | 0,006 | 0,006 | 0,002 | 27 | RMS |
| Wavefront Error (waves) | 1,413 | 1,200 | 0,350 | 25 | PV |
| | 0,261 | 0,257 | 0,047 | 18 | RMS |

### SDP – Case B

| Random | fwwe = 1 | p = 3,5 µm | a=0,5 | b=0,5 | d = 0 |
|---|---|---|---|---|---|
| Error type | Reference | Measured | Difference | Relative Difference (%) | |
| X-slopes (mrad) | 0,058 | 0,049 | 0,026 | 45 | PV |
| | 0,008 | 0,008 | 0,003 | 37 | RMS |
| Y-slopes (mrad) | 0,065 | 0,054 | 0,036 | 55 | PV |
| | 0,008 | 0,008 | 0,003 | 38 | RMS |
| Wavefront Error (waves) | 2,122 | 1,703 | 0,476 | 22 | PV |
| | 0,331 | 0,323 | 0,071 | 21 | RMS |

# APPENDIX C. STATIC PYRAMID TEST – DETAILED NUMERICAL RESULTS

In this Appendix the Table C1 presents the detailed data supporting the numerical simulations of the static pyramid test discussed in subsection 3.2, following the same format as in Appendix B.

Table C1: Detailed numerical results obtained for the static pyramid test.

### SFP - Case A

| Zern48 | fwwe = 1 | | a = 0 | b = 1 | d = 0 | |
|---|---|---|---|---|---|---|
| **Error type** | Reference | Measured | Difference | Relative Difference (%) | | |
| X-slopes (mrad) | 0,0308 | 0,0281 | 0,0244 | 79 | PV |
| | 0,0061 | 0,0061 | 0,0030 | 49 | RMS |
| Y-slopes (mrad) | 0,0308 | 0,0281 | 0,0244 | 79 | PV |
| | 0,0061 | 0,0061 | 0,0030 | 49 | RMS |
| Wavefront Error | 1,413 | 0,949 | 0,678 | 48 | PV |
| (waves) | 0,261 | 0,243 | 0,097 | 37 | RMS |

### SFP – Case B

| Random | fwwe = 1 | | a = 0 | b = 1 | d = 0 | |
|---|---|---|---|---|---|---|
| **Error type** | Reference | Measured | Difference | Relative Difference (%) | | |
| X-slopes (mrad) | 0,0580 | 0,0634 | 0,0428 | 74 | PV |
| | 0,0080 | 0,0080 | 0,0049 | 61 | RMS |
| Y-slopes (mrad) | 0,0651 | 0,0645 | 0,0512 | 79 | PV |
| | 0,0082 | 0,0082 | 0,0051 | 62 | RMS |
| Wavefront Error | 2,122 | 1,637 | 0,777 | 37 | PV |
| (waves) | 0,331 | 0,314 | 0,103 | 31 | RMS |

### SDP – Case A

| Zern48 | fwwe = 1 | p = 2 µm | a=0,5 | b=0,5 | d = 0 | |
|---|---|---|---|---|---|---|
| **Error type** | Reference | Measured | Difference | Relative Difference (%) | | |
| X-slopes (mrad) | 0,031 | 0,033 | 0,023 | 75 | PV |
| | 0,006 | 0,006 | 0,003 | 48 | RMS |
| Y-slopes (mrad) | 0,031 | 0,033 | 0,023 | 75 | PV |
| | 0,006 | 0,006 | 0,003 | 48 | RMS |
| Wavefront Error | 1,413 | 1,086 | 0,526 | 37 | PV |
| (waves) | 0,261 | 0,246 | 0,088 | 34 | RMS |

### SDP – Case B

| Random | fwwe = 1 | p = 3 µm | a=0,5 | b=0,5 | d = 0 | |
|---|---|---|---|---|---|---|
| **Error type** | Reference | Measured | Difference | Relative Difference (%) | | |
| X-slopes (mrad) | 0,058 | 0,067 | 0,042 | 72 | PV |
| | 0,008 | 0,008 | 0,005 | 58 | RMS |
| Y-slopes (mrad) | 0,065 | 0,071 | 0,051 | 78 | PV |
| | 0,008 | 0,008 | 0,005 | 59 | RMS |
| Wavefront Error | 2,122 | 1,670 | 0,681 | 32 | PV |
| (waves) | 0,331 | 0,318 | 0,091 | 27 | RMS |